\documentclass[12pt,english]{iopart}
\usepackage[T1]{fontenc}
\usepackage[latin1]{inputenc}
\usepackage{babel}
\usepackage{graphics}
\usepackage{latexsym}
\usepackage{iopams}

\begin{document}

\title{Nucleon and meson effective masses in the Relativistic Mean-Field
Theory }

\author{Ryszard Ma\'{n}ka \thanks{manka@us.edu.pl}
and Ilona Bednarek \thanks{bednarek@us.edu.pl}}

\address{Institute of Physics, Silesian University,
         Uniwersytecka 4, 40-007 Katowice, Poland.}

\begin{abstract}
Nucleon and meson effective masses in the nonlinear Relativistic
Mean-Field Theory (RMF) introducing a nonlinear \( \omega -\rho \) 
and \(\sigma \) coupling motivated by the Quark Meson Coupling model (QMC) 
is explored. It is shown that, in contrast to the usual
Walecka model, not only the effective nucleon mass \( m_{eff,N}
\) but also the effective \( \sigma ,\, \rho  \) meson masses \(
(m_{eff,\sigma },\, m_{eff,\rho }) \) and the effective \(
\omega  \) meson mass \( m_{eff,\omega } \) are nucleon density
dependent.
\end{abstract}


\submitto{\JPG}

\maketitle

\section{Introduction}

Relativistic mean-field theory seems to be a powerful tool in
describing various aspects of nuclear structure
\cite{Bog77,walecka,Rin96}. This theory provides an elegant and
economical framework, in which properties of nuclear matter
\cite{glen, weber}, finite nuclei and neutron star, as well as
the dynamics of heavy-ion collisions, can be calculated. Compared
to conventional nonrelativistic approaches, relativistic models
explicitly include mesonic degrees of freedom and describe
nucleons as Dirac quasiparticles. Moreover, the spin-orbit
interaction arises naturally from the Dirac-Lorenz structure of
the effective Lagrangian. Nucleons interact in a relativistic
covariant manner through the exchange of virtual mesons: the
isoscalar scalar \( \sigma  \)-meson, the isoscalar vector \(
\omega  \)-meson, and the isovector vector \( \rho  \)-meson. The
model is based on the one-boson exchange description of the
nucleon-nucleon interaction. One of the most interesting problems 
in intermediate energy physics which has been investigated with many 
approaches is an effective meson mass in nuclear
medium.

\section{Effective vector meson masses in nuclear medium.}

\begin{figure}

\caption{\label{masd} }

{\centering \resizebox*{10cm}{!}{\includegraphics{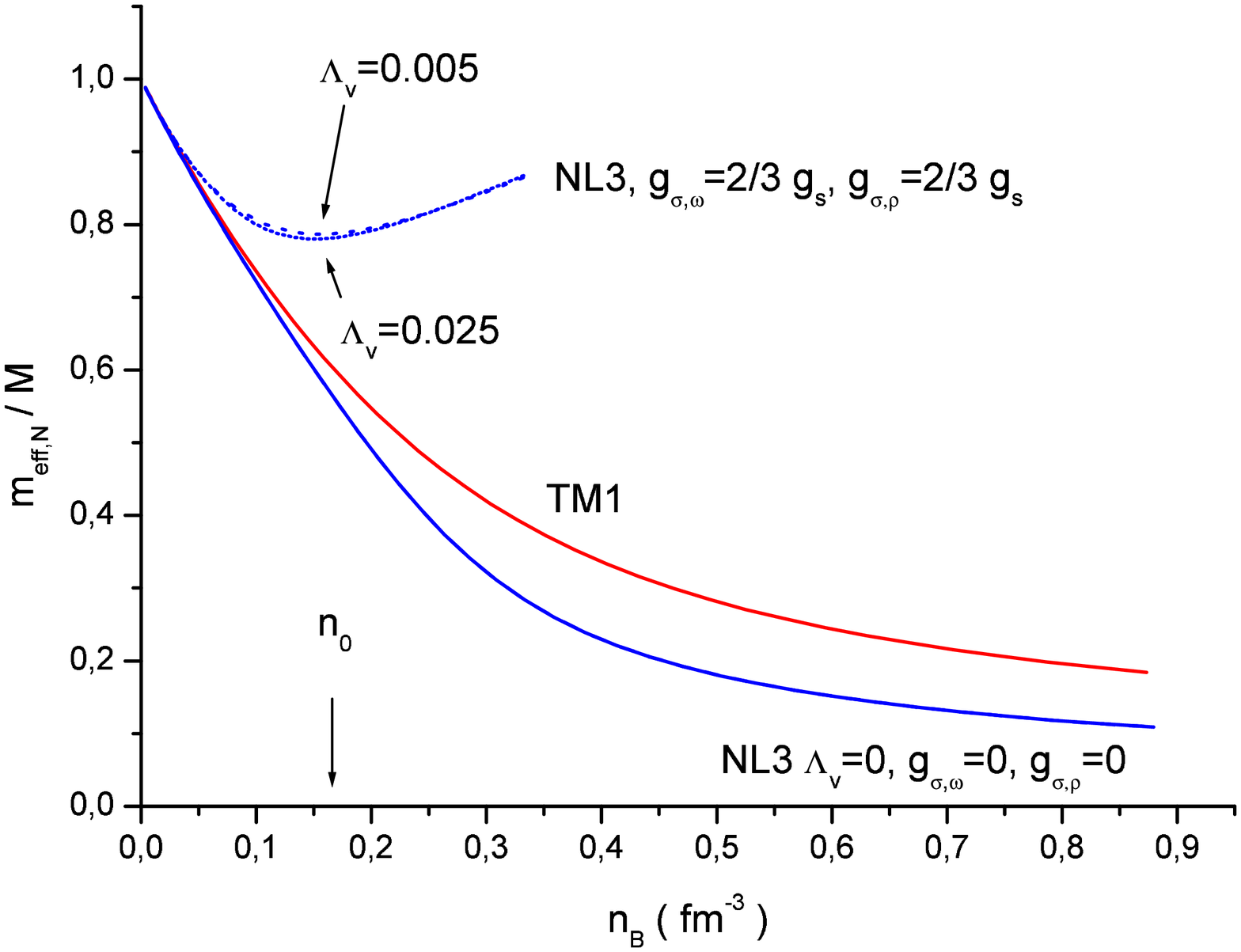}} \par}

Nucleon effective mass \protect\( m_{eff,N}=M-g_{N\sigma }\sigma \protect  \)
as the function of the baryon density \protect\( n_{B}\protect  \)
(\protect\( fm^{-3}\protect  \)).
\end{figure}

\begin{figure}
{\centering \resizebox*{10cm}{!}{\includegraphics{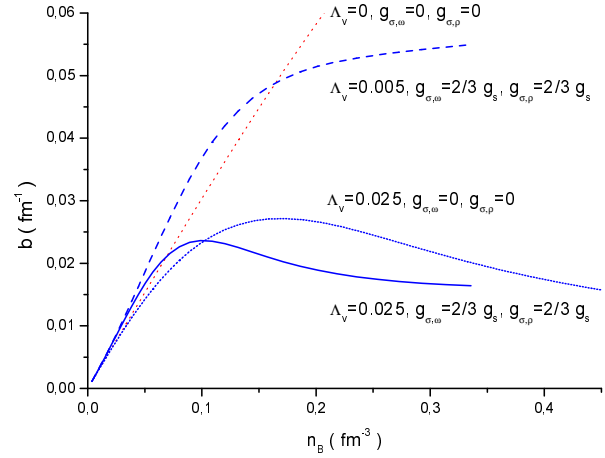}} \par}

\caption{\label{fig: rho}The static homogeneous solution for the \protect\protect\( \rho \protect \protect \)
meson field as the function of the baryon density \protect\protect\( n_{B}\protect \protect \)
(\protect\protect\( fm^{-3}\protect \protect \)).}
\end{figure}

In this paper the properties of nuclear matter in high-density and
neutron-rich regime are considered with the use of the relativistic
mean-field approximation. This approach implies the interaction of
nucleons through the exchange of meson fields, so the model considered
here comprises: nucleons and scalar, vector-isoscalar and vector-isovector
mesons. Consequently contributions coming from these components influence
the properties of nucleon. In this paper the nonlinear vector-isoscalar
self-interaction is dealt. Modification of this type was proposed
by Bodmer in order to achieve good agreement with the Dirac-Br\"{u}ckner
\cite{gmuca} calculations at high densities. The first version of
the \( \rho  \) meson field introduction is of a minimal type without
any nonlinearity and consists only of the coupling of this field to
nucleons. This case is enlarged by the nonlinear vector-isoscalar
and vector-isovector interaction and the vector-scalar interactions
which modify the density dependence of the \( \rho  \) mean field
and the energy symmetry. Such an extension was inspired by the paper
\cite{piek} in which the authors indicate the existence of a relation
between the neutron-rich skin of a heavy nucleus and the properties
of a neutron star crust. In the quark meson coupling model \cite{qmc1} nucleon 
properties are modified by the meson coupling directly to  the quarks 
and not to the nucleons ( paper by Li et.al. \cite{qli}). The
RMF theory is an effective one, as coupling constants are determined
by the bulk properties of nuclear matter such as saturation density,
binding energy, compressibility and the symmetry energy.\\
 The starting point in this paper is the construction of the effective
Lagrangian \cite{hir1,Ser86, Ser92} intended for application to the system
described above. The scalar \( \sigma  \), isoscalar-vector \( \omega  \)
and isovector-vector \( \rho  \) mesons are denoted by \( \varphi  \),
\( \omega _{\mu } \), \( b_{\mu }^{a} \), respectively.
The Lagrange density function for this model has the following form
\begin{eqnarray}
{\mathcal{L}} & =\frac{1}{2}\partial _{\mu }\varphi \partial ^{\mu }\varphi -U(\varphi )-\frac{1}{4}\Omega _{\mu \nu }\Omega ^{\mu \nu }+\frac{1}{2}(M_{\omega }-g_{\omega \sigma }\varphi )^{2}\omega _{\mu }\omega ^{\mu }+\frac{1}{4}c_{3}(\omega _{\mu }\omega ^{\mu })^{2}& \nonumber \label{lag} \\
 & -\frac{1}{4}R_{\mu \nu }^{a}R^{a\mu \nu }+\frac{1}{2}(M_{\rho }-g_{\rho \sigma }\varphi )^{2}b^{a}_{\mu }b^{a\mu }+\Lambda _{v}(g_{\rho }g_{\omega })^{2}(\omega _{\mu }\omega ^{\mu })(b^{a}_{\mu }b^{a\mu }) & \nonumber \\
 & -\frac{1}{4}F_{\mu \nu }F^{\mu \nu }+i\overline{\psi }\gamma ^{\mu }D_{\mu }\psi -\overline{\psi }(M-g_{N\sigma }\varphi )\psi .
\end{eqnarray}
The field tensors \( R_{\mu \nu }^{a} \), \( \Omega _{\mu \nu } \),
\( F_{\mu \nu } \) and the covariant derivative \( D_{\mu } \) are
given by
\begin{equation}
R_{\mu \nu }^{a}=\partial _{\mu }b^{a}_{\nu }-\partial _{\nu }b^{a}_{\mu }+g_{\rho }\varepsilon _{abc}b_{\mu }^{b}b_{\nu }^{c} ,
\end{equation}
\begin{equation}
\Omega _{\mu \nu }=\partial _{\mu }\omega _{\nu }-\partial _{\nu }\omega _{\mu } ,
\end{equation}
\begin{equation}
F_{\mu \nu }=\partial _{\mu }A_{\nu }-\partial _{\nu }A_{\mu } ,
\end{equation}
\begin{equation}
D_{\mu }=\partial _{\mu }+\frac{1}{2}ig_{\rho }b^{a}_{\mu }\tau ^{a}+ig_{\omega }\omega _{\mu }+iQ_{e}A_{\mu } .
\end{equation}
The potential function \( U(\varphi ) \) possesses a polynomial
form introduced by Boguta and Bodmer \cite{bodmer} in order to get
a correct value of the compressibility \( K \) of nuclear matter
at saturation density
\begin{equation}
U(\varphi )=\frac{1}{2}m^{2}_{s}\varphi ^{2}+\frac{1}{3}g_{2}\varphi ^{3}+\frac{1}{4}g_{3}\varphi ^{4}.
\end{equation}
The nucleon field \( \psi  \) has a form of a column matrix composed
of proton and neutron fields, respectively
\begin{equation}
\psi =\left( \begin{array}{l}
\psi _{n}\\
\psi _{p}
\end{array}\right).
\end{equation}
The nucleon mass is denoted by \( M \) whereas \( m_{s} \), \( M_{\omega } \),
\( M_{\rho } \) are masses assigned to the meson fields. The parameters
entering the Lagrangian function (\ref{lag}) are the coupling constants
\( c_{3} \), \( g_{\omega } \), \( g_{\rho } \), \( g_{N\sigma } \)
(Table \ref{tab1}) for meson fields, \( e^{2}/4\pi =1/137 \) for
photon field and the self-interacting coupling constants \( g_{2} \)
and \( g_{3} \). The coupling constants \( g_{\omega \sigma } \)
and \( g_{\rho \sigma } \) in the QMC model \cite{qmc2} are related to the constant \( g_{N\sigma } \)
by the relations
\begin{equation}
g_{\omega \sigma }=\frac{2}{3}g_{N\sigma },\, \, g_{\rho \sigma }=\frac{2}{3}g_{N\sigma } .
\end{equation}
The Lagrangian function (\ref{lag}) includes also nonlinear terms
\begin{eqnarray}
 &  & \frac{1}{4}c_{3}(\omega _{\mu }\omega ^{\mu })^{2}+\Lambda _{v}(g_{\rho }g_{\omega })^{2}(\omega _{\mu }\omega ^{\mu })(b^{a}_{\mu }b^{a\mu })\\
 & - & g_{\omega \sigma}M_{\omega }\varphi \, \omega _{\mu }\omega ^{\mu }+\frac{1}{2}g_{\omega \sigma}^{2}\varphi ^{2}\omega _{\mu }\omega ^{\mu }-g_{\rho \sigma}M_{\omega }\varphi \, b^{a}_{\mu }b^{a\mu }+\frac{1}{2}g_{\rho \sigma}^{2}\varphi ^{2}b^{a}_{\mu }b^{a\mu }\nonumber
\end{eqnarray}
which affect remarkably properties of nucleons. The first term was
proposed by Bodmer in order to achieve good agreement with the
Dirac-Br\"{u}ckner \cite{gmuca} calculations at high densities.
All terms which involve
the \( b_{\mu }^{a} \) field change the density dependence of the
symmetry energy and cause the change in the density dependence of
the \( \rho  \) field. The modification of the density dependence
of the \( \rho  \) mean field (\ref{fig: rho}) can lead to the change
of the neutron-skin thickness \cite{piek}. This effect is observed
in the Parity Radius Experiment (PREX) at the Jefferson Laboratory
which aims to measure the neutron radius in \( ^{208} \)Pb via parity
violating electron scattering \cite{prex}.\\
The field equations for meson fields \( \Phi _{A}=\{ \)\( \varphi  \),
\( \omega _{\mu } \), \( b_{\mu }^{a}\} \) are
\begin{eqnarray}
\Box \varphi  & = & m_{s}^{2}\varphi +g_{2}\varphi ^{2}+g_{3}\varphi ^{3}-g_{\sigma N}\overline{\psi }\psi +(M_{\omega }-g_{\omega \sigma }\varphi )g_{\omega \sigma }\omega _{\mu }\omega ^{\mu }\label{egg1} \\
 & + & (M_{\rho }-g_{\rho \sigma }\varphi )g_{\rho \sigma }b_{\mu }^{a}b^{a\mu } , \nonumber
\end{eqnarray}
\begin{eqnarray}
-\partial _{\mu }\Omega ^{\mu \nu } & = & (M_{\omega }-g_{\omega \sigma }\varphi )^{2}\omega ^{\nu }+c_{3}(\omega _{\mu }\omega ^{\mu })\omega ^{\nu } \label{egg3} \\
 & + & 2\Lambda _{v}(g_{\omega }g_{\rho })^{2}(b^{a}_{\mu }b^{a\mu })\omega ^{\nu }-g_{\omega }J_{B}^{\nu } , \nonumber
\end{eqnarray}
\begin{equation}
\label{egg2}
-D_{\mu }R^{\mu \nu a}=(M_{\rho }-g_{\rho \sigma }\varphi )^{2}b^{\nu a}+2\Lambda _{v}(g_{\omega }g_{\rho })^{2}(\omega _{\mu }\omega ^{\mu })b^{a\nu }-g_{\rho }\delta _{a,3}J_{3}^{\nu }.
\end{equation}
Sources that appear in these equations are the baryon current
\begin{equation}
J_{B}^{\nu }=\overline{\psi }\gamma ^{\nu }\psi
\end{equation}
and the isospin current which exists only in the asymmetric matter
\begin{equation}
J^{\nu }_{3}=\frac{1}{2}\overline{\psi }\gamma ^{\nu }\tau ^{3}\psi
\end{equation}
The Maxwell equations are of the form
\[
-\partial _{\mu }F^{\mu \nu }=-e\, J^{\nu }
\]
with the electric current given by
\begin{equation}
J^{\nu }=\frac{1}{2}\overline{\psi }\gamma ^{\nu }Q_{e}\psi .
\end{equation}
The timelike components of the baryon and isospin currants are the
baryon density \( \rho _{B}=J^{0}_{B} \) and the isospin density
\( \rho _{a}=J^{0}_{a} \). The conserved baryon and isospin charges
are given by the following relations:
\[
Q_{B}=\int d^{3}x\psi ^{+}\psi =\int d^{3}x(\psi _{n}^{+}\psi _{n}+\psi _{p}^{+}\psi _{p}),\]
\[
Q_{3}=\frac{1}{2}\int d^{3}x\psi ^{+}\sigma ^{3}\psi =\frac{1}{2}\int d^{3}x(\psi _{n}^{+}\psi _{n}-\psi _{p}^{+}\psi _{p}) ,
\]
The electric charge is defined by
\[
Q=\int d^{3}x\psi ^{+}Q_{e}\psi =\int d^{3}x\psi _{p}^{+}\psi _{p}\, \, \, \, ,Q_{e}=\frac{1}{2}(I-\tau ^{3})=\left(
\begin{array}{cc}
0 & 0\\
0 & 1
\end{array}\right).
\]
The Dirac equations for nucleons that are obtained from the Lagrangian
function have the following form
\begin{equation}
\label{egg4}
i\gamma ^{\mu }D_{\mu }\psi -(M-g_{N\sigma }\varphi )\psi =0.
\end{equation}
The physical system is defined by the thermodynamic potential \cite{fet}
\begin{equation}
\Omega =-kTlnTr(e^{-\beta (H-\mu Q_{B}-\mu _{3}Q_{3})})
\end{equation}
where H stands for the Hamiltonian and is defined as
\begin{equation}
H=\sum _{A}\int d^{3}x\{\partial _{0}\Phi _{A}\pi ^{A}-\mathcal{L}\}.
\end{equation}
\( \pi ^{A}=\frac{\partial \mathcal{L}}{\partial (\partial _{0}\Phi _{A})} \)
are momenta connected with fields \( \Phi _{A}=\{ \)\( \varphi  \),
\( \omega _{\mu } \), \( b_{\mu } \),\( \psi \} \) which denote
all fields considered in this model. The averaged charges can be obtained
from the thermodynamic potential with the use of the relations
\begin{equation}
\frac{\partial \Omega }{\partial \mu }=-<Q_{B}>,\, \, \, \, \frac{\partial \Omega }{\partial \mu _{3}}=-<Q_{3}>.
\end{equation}
In this paper the variational method based on the Feynman-Bogolubov
inequality \cite{rm} is incorporated (see more details in \cite{rm1})
\begin{equation}
\Omega \leq \Omega _{1}=\Omega _{0}(m_{eff})+<H-H_{0}>_{0} .
\end{equation}
\( \Omega _{0} \) is the thermodynamic potential of the trial system
which as effectively free quasiparticle system is described by the
Lagrange function
\begin{eqnarray}
 & {\mathcal{L}}_{0}(m_{eff})=\frac{1}{2}\partial _{\mu }\overline{\varphi }\partial ^{\mu }\overline{\varphi }-\frac{1}{2}m_{eff,\sigma }^{2}\overline{\varphi }^{2}-\frac{1}{4}\overline{\Omega }_{\mu \nu }\overline{\Omega }^{\mu \nu }+\frac{1}{2}m_{eff,\omega }^{2}\overline{\omega }_{\mu }\overline{\omega }^{\mu } & \nonumber \\
 & -\frac{1}{4}\overline{G}_{\mu \nu }^{a}\overline{G}^{a\mu \nu }+\frac{1}{2}\sum _{a}m^{2}_{eff,\rho ,a}\overline{b}_{\mu }^{a}\overline{b}^{a\mu } & \\
 & +\overline{\psi }(i\gamma ^{\mu }\overline{D}_{\mu }-m_{eff,N})\psi  & \nonumber
\end{eqnarray}
with the field tensor now defined by
\[
\overline{G}_{\mu \nu }^{a}=\partial _{\mu }\overline{b}^{a}_{\nu }-\partial _{\nu }\overline{b}^{a}_{\mu }
\]
and
\[
\overline{\Omega }_{\mu \nu }=\partial _{\mu }\overline{\omega }_{\nu }-\partial _{\nu }\overline{\omega }_{\mu }.
\]
The fields \( \Phi _{A} \) can be written as a sum of two components,
the effectively free quasiparticle fields \( \overline{\Phi }_{A}={\overline{\varphi },\overline{\omega }_{\mu },\overline{b}^{a}_{\mu }} \)
and the classical boson condensates \( \xi _{A}={\sigma ,w,b} \)
\begin{equation}
\Phi _{A}=\overline{\Phi }_{A}+\xi _{A},
\end{equation}
which in the case of particular meson fields results in
\begin{equation}
\varphi =\overline{\varphi }+\sigma
\end{equation}
\begin{equation}
\omega _{\mu }=\overline{\omega }_{\mu }+w_{\mu },\, \, w_{\mu }=\delta _{\mu ,0}w
\end{equation}
\begin{equation}
\rho _{\mu }^{a}=\overline{b}^{a}_{\mu }+\beta ^{a}_{\mu },\, \, \beta ^{a}_{\mu }=\delta ^{a,3}\delta _{\mu ,0}\, b
\end{equation}
Both the boson and fermion masses \( (m_{B},m_{F}) \) as well as
the \( \xi _{A}=\{\sigma ,\, w,\, b\} \) fields are treated as the
variational parameters in the effective potential. The covariant derivative
for the trial system is
\begin{equation}
\overline{D}_{\mu }=\partial _{\mu }+\frac{1}{2}ig_{\rho }\beta ^{a}_{\mu }\tau ^{a}+ig_{\omega }w_{\mu }
\end{equation}
This introduces the fermion interaction with homogenous boson condensate
\( w_{\mu },\, \beta ^{a}_{\mu }. \) The fermion quasiparticle obeys
the Dirac equation
\begin{equation}
(i\gamma ^{\mu }\, \overline{D}_{\mu }-m_{eff,N})\psi =0
\end{equation}
The masses entering the Lagrangian function \( L_{0}(m_{eff}) \)
of the trial system calculated from the extremum conditions
\[
\frac{\partial \Omega _{1}}{\partial m^{2}_{eff,\sigma }}=0,\, \, \, \frac{\partial \Omega _{1}}{\partial m^{2}_{eff,\omega }}=0,\, \, \, \frac{\partial \Omega _{1}}{\partial m^{2}_{eff,\rho }}=0,\, \, \, \frac{\partial \Omega _{1}}{\partial m_{eff,N}}=0
\]
now are the effective once. They are given by the following equations
\begin{eqnarray}
m^{2}_{eff,\sigma } & = & m^{2}_{s}+3g_{3}<\overline{\varphi }^{2}>_{0}+2b_{2}\sigma +3b_{3}\sigma ^{2}\label{mb} \\
 & - & g^{2}_{\omega \sigma }(w^{2}+<\overline{\omega }_{\mu }\overline{\omega }^{\mu }>_{0})-g^{2}_{\rho \sigma }(b^{2}+<\overline{b}^{a}_{\mu }\overline{b}^{a\mu }>_{0}), \nonumber
\end{eqnarray}
\begin{eqnarray}
m^{2}_{eff,\omega } & = & (M_{\omega }-g_{\sigma \omega }\sigma )^{2}+g_{\omega \sigma }^{2}<\overline{\varphi }^{2}>_{0}+c_{3}(w^{2}+<\overline{\omega }_{\mu }\overline{\omega }^{\mu }>_{0})\label{mw} \\
 & + & 2\Lambda _{v}(g_{\rho }g_{\omega })^{2}(b^{2}+<\overline{b}^{a}_{\mu }\overline{b}^{a\mu }>_{0})\nonumber
\end{eqnarray}
\begin{eqnarray}
 & m^{2}_{eff,\rho ,a}=(M_{\rho }-g_{\rho \sigma }\sigma )^{2}+g_{\rho \sigma }^{2}<\overline{\varphi }^{2}>_{0}+2(g_{\rho }g_{\omega })^{2}\Lambda _{v}(w^{2}+<\overline{\omega }_{\mu }\overline{\omega }^{\mu }>_{0}) & \nonumber \\
 & -g_{\rho }^{2} b^{2}(1-\delta _{a,3})-\frac{3}{2}g_{\rho }^{2}\sum _{c\neq a}<\overline{b}^{c}_{\mu }\overline{b}^{c\mu }>_{0}^{2} & \label{mr} \\
 & \, \, \, \, ^{\longrightarrow }_{T=0}\, \, (M_{\rho }-g_{\rho \sigma }\sigma )^{2}-g_{\rho }^{2}b^{2}(1-\delta _{a,3})+2(g_{\rho }g_{\omega })^{2}\Lambda _{v}w^{2} & \nonumber
\end{eqnarray}
\begin{equation}
\label{mf}
m_{eff,N}=M-g_{N\sigma }\sigma .
\end{equation}
In the presence of the \( \rho  \) field condensation \( b \) the
\( \rho  \) mass \( m_{eff,\rho ,a} \) is splitting. This is a result
of the 'nonabelian' character of the \( SU(2) \) symmetry which is
explicitely broken by the \( M_{\rho } \) term. In the homogeneous
case the free energy reaches the minimum at \( \sigma  \). The nucleon
effective mass \( m_{eff,N} \) dependence on the baryon density \( n_{B} \)
is presented in Fig. \ref{masd}. Results are shown for the TM1 and
NL3 parameter sets without additional coupling and for NL3 supplemented
by the vector-scalar coupling.

\begin{figure}
{\centering \resizebox*{10cm}{!}{\includegraphics{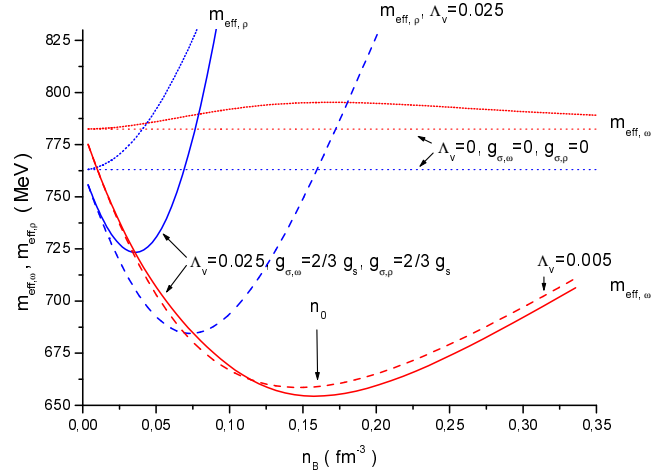}} \par}

\caption{\label{mases}The mesons \protect\protect\( \omega \protect \protect \)
and \protect\protect\( \rho \protect \protect \) effective mass as
the function of the baryon density \protect\protect\( n_{B}\protect \protect \)
(\protect\protect\( fm^{-3}\protect \protect \)).}
\end{figure}

\begin{figure}
{\centering \resizebox*{10cm}{!}{\includegraphics{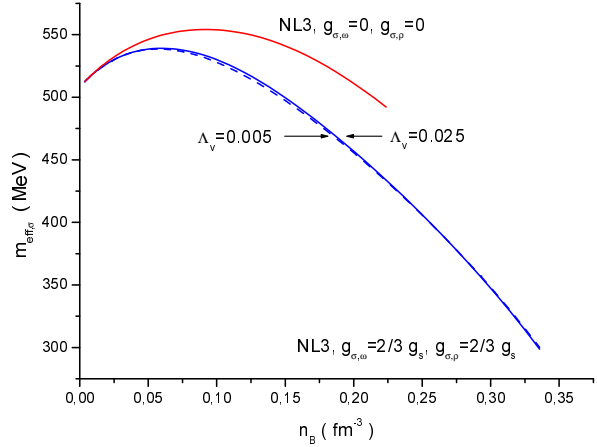}} \par}

\caption{\label{msigma}The mesons \protect\protect\( \omega \protect \protect \)
and \protect\protect\( \sigma \protect \)\protect\( \protect \protect \)
effective mass as the function of the baryon density \protect\protect\( n_{B}\protect \protect \)
(\protect\protect\( fm^{-3}\protect \protect \)).}
\end{figure}

The meson effective masses as functions of the baryon density \( n_{B} \)
are presented in Figs. \ref{mases} and \ref{msigma}. This figure
provides a comparison of the meson effective masses obtained for different
values of the coupling constants. Straight dotted lines show the \( \omega  \)
and \( \rho  \) masses in the simplest model with no additional nonlinear
terms. When this model is enlarged by the nonlinear vector-isoscalar
and vector-isovector and the vector-scalar interactions the effective
masses are obtained. As baryon number increases the \( \omega  \)
meson effective mass increases and then become smaller. The \( \rho  \)
meson effective mass increases with the increasing baryon number.

The symmetry energy density (for \( T=0 \)) \( \varepsilon _{s}=E_{s}/n_{B} \)
can be evaluated taking into account the contributions of the \( \rho  \)
meson field to the energy of the system
\begin{equation}
E_{s}=\frac{1}{2}m_{eff,\rho }^{2}b^{2}=\frac{1}{2}(M_{\rho }-g_{\rho \sigma }\sigma )^{2}\, b^{2}+\Lambda _{v}(g_{\rho }g_{\omega })^{2}w^{2}b^{2}
\end{equation}
\begin{figure}
{\centering \resizebox*{10cm}{!}{\includegraphics{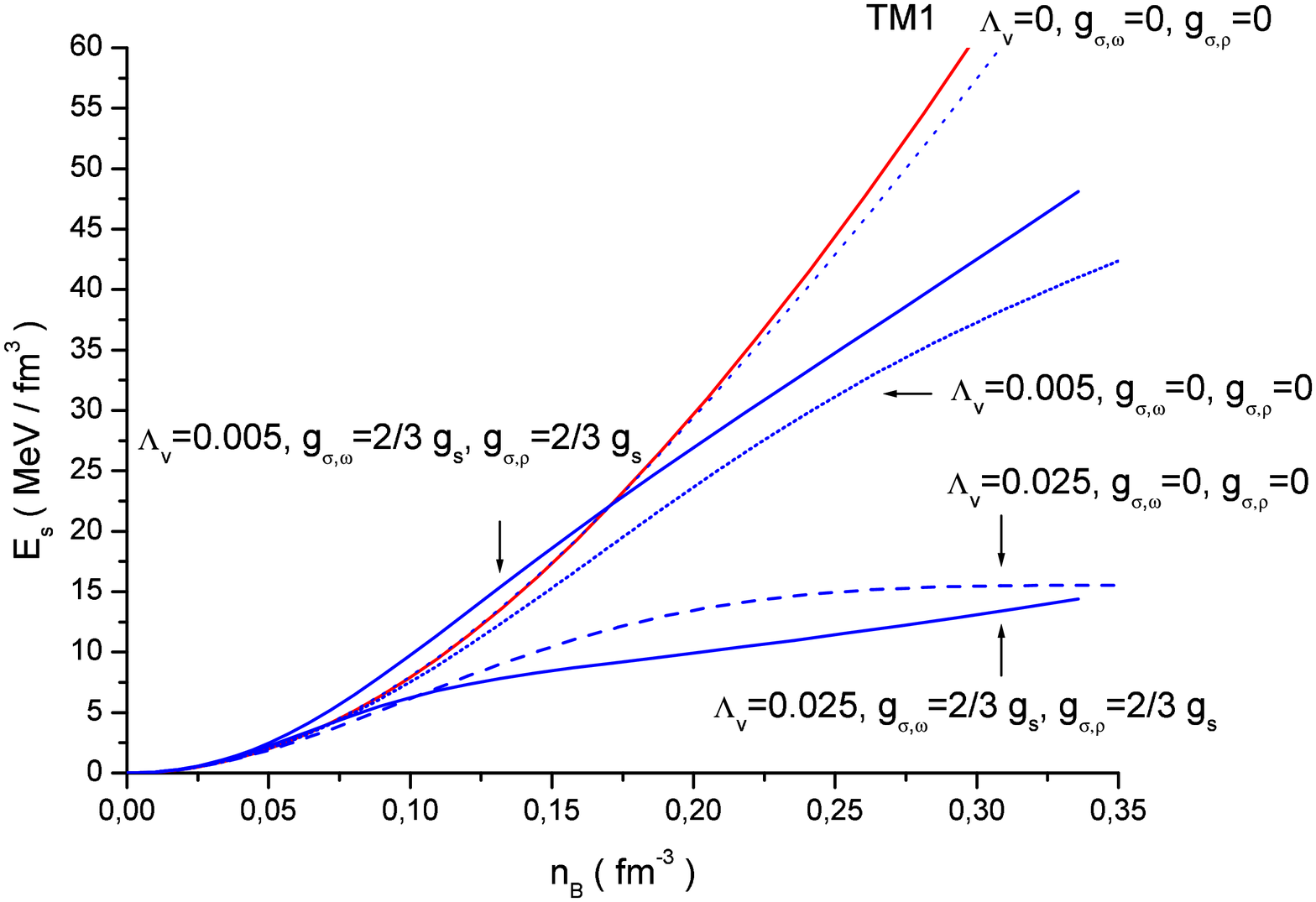}} \par}

\caption{\label{rhof}The symmetry energy \protect\protect\( E_{s}\protect \protect \)
as the function of the baryon density \protect\protect\( n_{B}\protect \protect \)
(\protect\protect\( fm^{-3}\protect \protect \)).}
\end{figure}
Using the equation of motion the static homogeneous solution for the
\( \rho  \) field takes the following form
\begin{eqnarray}
b=\frac{g_{\rho }}{m_{eff,\rho }^{2}}Q_{3}=\frac{g_{\rho }n_{B}}{2m_{eff,\rho }^{2}}\delta  &  & \label{symmetry} \\
\delta =(n_{n}-n_{p})/n_{B} &  & \label{symmetry2}
\end{eqnarray}
and is presented in Fig. \ref{fig: rho}.
Taking into account the form of the equations (\ref{symmetry}) and
(\ref{symmetry2}) the symmetry energy now is given by
\[
\varepsilon _{s}=\frac{g_{\rho }^{2}n_{B}}{8m_{eff,\rho }^{2}}\delta ^{2}\, \, \, \rightarrow \, \, \, \frac{g_{\rho }^{2}n_{B}}{8M_{\rho }^{2}}\delta ^{2} .
\]
In the case of no additional coupling constants the result reproduces
that of the simplest model without any nonlinearities. Including the
\( \omega -\rho  \) and the vector-scalar coupling the \( \varepsilon _{s} \)
is no longer linear with baryon density \( n_{B} \). Including the
\( \omega -\rho  \) and the vector-scalar coupling the \( \varepsilon _{s} \)
is no longer linear with baryon number density \( n_{B} \). For the
purpose of the present study, we choose two basic RMF parameterizations
(Table \ref{tab1}): NL3 \cite{Lal97}, TM1 \cite{tm1} and one based on QMC
approach \cite{qmc2}. All parameters are chosen in such a form to reproduce 
the nucleon symmetric phase properties (the binding energy \( E_{0} \),
the Fermi momentum, the density and the compresibility \( K \)) at
the satutation point (Table \ref{tab3}).
\begin{table}%
\begin{center}
\begin{tabular}{|c||c|c|c|}
\hline 
Parameter&
NL3&
TM1&
QMC\\
\hline
\hline 
\( E_{0} \)~(MeV)&
-15.92&
-16.26&
-16.65\\
\hline 
\( k_{F,0} \)~(\( fm^{-1}) \)&
1.24&
1.29&
1.297\\
\hline 
\( \rho _{0} \)~\( (fm^{-3}) \)&
0.14&
0.145&
0.147\\
\hline 
\( K \)~(MeV)&
170.24&
281.53&
284.71\\
\hline
\end{tabular}
\end{center}
\caption{\label{tab3}
The nucleon symmetric phase properties (the binding energy \( E_{0} \),
the Fermi momentum, the density and the compresibility \( K \)) at
the saturation point.}
\end{table}%

\begin{figure}
{\centering \resizebox*{10cm}{!}{\includegraphics{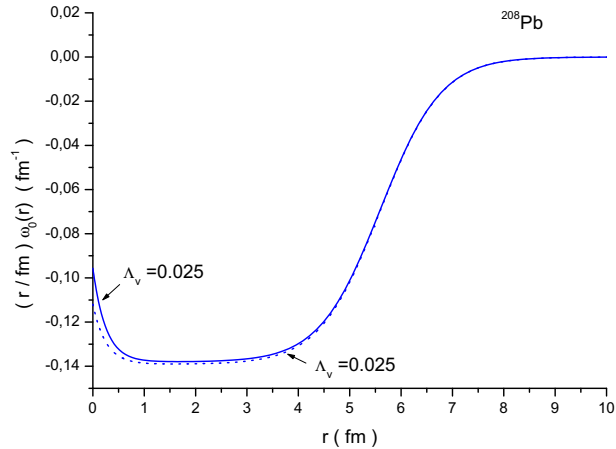}} \par}

\caption{\label{omo}Profile of the \protect\protect\( \omega \protect \protect \)
meson field inside the \protect\protect\( ^{208}\protect \protect \)Pb
nucleus (\protect\( g_{\omega \sigma }=g_{\rho \sigma }=0\protect \)
case).}
\end{figure}

\begin{figure}
{\centering \resizebox*{10cm}{!}{\includegraphics{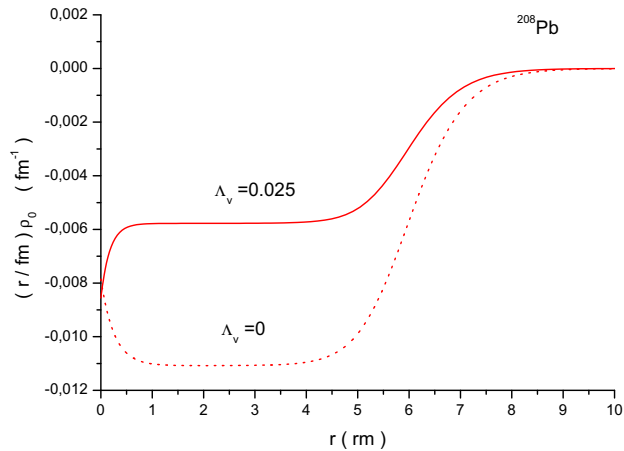}} \par}

\caption{\label{rhor}Profile of the \protect\protect\( \rho \protect \protect \)
meson field inside the \protect\protect\( ^{208}\protect \protect \)Pb
nucleus (\protect\( g_{\omega \sigma }=g_{\rho \sigma }=0\protect \)
case).}
\end{figure}

The case of static nonhomogeneous nuclear matter configuration can
be described by the Euler-Lagrange equations (\ref{egg1}, \ref{egg2},
\ref{egg3}, \ref{egg4}) taking the quantum average of these equations
with respect to the effective quasiparticle system \( H_{0} \) which
in turn is obtained with the use of the effective quasiparticle Lagrangian
\( {\mathcal{L}}_{0}(m_{eff}) \). The results for a spherically symmetric
system become
\begin{eqnarray}
\Delta \sigma (r) & = & \frac{\partial U_{eff}}{\partial \sigma } . \label{eqr1}
\end{eqnarray}
\begin{equation}
\label{eqr2}
\Delta w(r)=m^{2}_{eff,\omega }\, w(r)-g_{\omega }\, Q_{B},
\end{equation}
\begin{equation}
\label{eqr3}
\Delta b(r)=m^{2}_{eff,\rho }\, b(r)-g_{\rho }\, Q_{3}
\end{equation}
where the effective potential for the scalar meson is given by
\begin{eqnarray}
 & U_{eff}=(g_{2}<\overline{\varphi }^{2}>_{0}+g_{\omega \sigma }M_{\omega }(w^{2}+<\overline{\omega }_{\mu }\overline{\omega }^{\mu }>_{0})+g_{\rho \sigma }M_{\rho }(b^{2}+<\overline{b}^{a}_{\mu }\overline{b}^{a\mu }>_{0}))\sigma  & \nonumber \\
 & +\frac{1}{2}(m^{2}_{s}+3g_{3}<\overline{\varphi }^{2}>_{0}-g^{2}_{\omega \sigma }(w^{2}+<\overline{\omega }_{\mu }\overline{\omega }^{\mu }>_{0})-g^{2}_{\rho \sigma }(b^{2}+<\overline{b}^{a}_{\mu }\overline{b}^{a\mu }>_{0}))\, \sigma ^{2} & \nonumber \\
 & +\frac{1}{3}g_{2}\sigma ^{3}+\frac{1}{4}g_{3}\sigma ^{4}\label{ueff} .
\end{eqnarray}
The construction of \( U_{eff} \) is such that
\[
m^{2}_{eff,\sigma }=\frac{\partial ^{2}U_{eff}}{\partial \sigma ^{2}}.
\]
The \( \rho  \) condensate has a mass (\( m_{eff,\rho }=m_{eff,\rho ,3} \)
at \( T=0 \))
\begin{eqnarray}
 & m^{2}_{eff,\rho }=(M_{\rho }-g_{\rho \sigma }\sigma )^{2}+g_{\rho \sigma }^{2}<\overline{\varphi }^{2}>_{0}+2(g_{\rho }g_{\omega })^{2}\Lambda _{v}(w^{2}+<\overline{\omega }_{\mu }\overline{\omega }^{\mu }>_{0}) & \nonumber \\
 & -g_{\rho }^{2}\sum _{c\neq 3}<\overline{b}^{c}_{\mu }\overline{b}^{c\mu }>_{0}^{2} . &
\end{eqnarray}
In the obtained results the thermal contributions from meson fields
are present, however, they are not involved in calculations since
meson masses are too large to be relevant. The Dirac equation for
nucleons (i=p,n) has the following form
\begin{equation}
\label{eqr4}
\{-i\alpha \nabla +V(r)+\beta [M-S(r)]\}\psi _{i}=\epsilon _{i}\psi _{i}
\end{equation}
with the potentials:
\[
V(r)=g_{\omega }\, w(r)+g_{\rho }\, \sigma ^{3}b(r)+eQ_{e}A_{0}(r),
\]
\[
S(r)=g_{N\sigma }\sigma (r).
\]
In the non-honomegenous nuclear matter distribution the effective
masses are space dependent and act rather like external
potentials. The self-consistent equations (\ref{eqr1}, \ref{eqr2},
\ref{eqr3}, \ref{eqr4}), namely the Dirac equation with potential
terms for the nucleons and the Klein-Gordon type equations with
sources for the mesons and the photon are the base for the
numerical description of nuclei \cite{ringc}. The simplified
solutions (cf. Fig. \ref{omo} and Fig.\ref{rhof}) can be obtained
assuming the Fermi shape for the neutron and proton density
distribution \cite{piek}. Spherically symmetric solutions for the
\( \omega  \) and \( \rho  \) meson fields for the \( ^{208} \)Pb
nucleus (\( g_{\omega \sigma }=g_{\rho \sigma }=0 \) case) are
presented in Fig. \ref{omo} and Fig. \ref{rhor}.
More realistic calculations, which will be the subject of future investigations,
can be obtained with the use of the self-consistent Thomas-Fermi approximation
\cite{prov1}, \cite{prov2}. In the case of the \( \omega  \) meson
the influence of the \( \omega -\rho  \) coupling is rather small
whereas there exists a marked difference in the \( \rho  \) meson
field profile when the \( \omega -\rho  \) coupling is present (cf.
(Fig. \ref{rhor})).

\begin{table}
\begin{tabular}{|c|c|c|c|c|c|}
\hline
\( g_{\rho } \)\cite{piek}&
 \( \Lambda _{v} \)\cite{piek}&
 \( R_{p}\, (fm) \)\cite{piek}&
 \( (R_{n}-R_{p})\, (fm) \)\cite{piek}&
 \( m_{eff,\omega }\, (MeV) \)&
 \( m_{eff,\rho }\, (MeV) \)\\
\hline
\( 8.92 \)&
 \( 0 \)&
 \( 5.460 \)&
 \( 0.280 \)&
 \( 782.5 \)&
 \( 763.0 \)\\
\hline
\( 10.75 \)&
 \( 0.025 \)&
 \( 5.469 \)&
 \( 0.209 \)&
 \( 785.4 \)&
 \( 1141.6 \) \\
\hline
\end{tabular}

\par{}

\caption{\label{tab2}The meson effective masses inside the \protect\protect\( ^{208}\protect \protect \)Pb
nucleus (\protect\( g_{\omega \sigma }=g_{\rho \sigma }=0\protect \)
case).}
\end{table}

\section{Conclusions }
The model presented here includes the original Walecka model with
the nucleon fields and the kinetic and mass terms for scalar and
vector mesons. Scalar nonlinear terms were proposed by Boguta and
Bodmer. Boguta and Price introduced quartic vector
self-interaction. The chosen form of the Lagrange function which
was supplemented by the additional omega-rho coupling was inspired
by the work by Horowitz and Piekarewicz \cite{piek} and by the QMC model 
\cite{qmc2}. 
This additional nonlinear meson interaction can be explained in terms
of effective meson masses that are modified in the nuclear medium.
The additional nonlinear couplings $\Lambda_{V}$ and $g_{\omega
\sigma}$ and $g_{\rho \sigma}$ change the density dependence of
the omega and rho fields, whereas $\Lambda _{V}$ itself influences
the density dependence of the symmetry energy. Taking into account
these additional  interaction terms the selfconsistent equations
for vector mesons were obtained. The nucleon and scalar
self-interaction generates different from zero expectation value
of the scalar field and the effective nucleon mass \( m_{eff,N}
\). The mutual interactions of vector meson fields give as a
result the effective meson masses \( m_{eff,\omega } \) and \(
m_{eff,\rho } \). The appearance of the effective meson masses
changes the energy symmetry of the system and has an influence on
the behaviour of the \( \rho  \) field. All calculations were
performed with the use of NL3 \cite{Lal97} and TM1 \cite{tm1}
parameter sets. They not only correctly reproduce nuclear
properties but also are consistent with the idea of naturalness as
well. The force of NL3 stems from the fit including exotic nuclei,
neutron radii, and information on giant resonances. The parameter
set named TM1 has been applied to deformed nuclei, triaxal
deformations, nuclear excitations \cite{hir1, hir2, ma}. The last
column mark by QMC represents parameters produced by the Quark
Meson Coupling model (QMC) \cite{qmc1,qmc2}.
\begin{table}%
\begin{tabular}{|c|c|c|c|c|c|}
\hline 
Parameter&
 NL3 \cite{Lal97}&
 NL3 \cite{piek}&
 NL3 \cite{piek}&
 TM1 \cite{tm1}&
 QMC \cite{qmc2}\\
\hline
\( M \)&
 \( 938\, MeV \)&
 \( 938\, MeV \)&
 \( 938\, MeV \)&
 \( 938\, MeV \)&
 \( 939\, MeV \)\\
\hline
\( M_{w} \)&
 \( 795.4\, MeV \)&
 \( 795.4\, MeV \)&
 \( 795.4\, MeV \)&
 \( 783\, MeV \)&
 \( 783\, MeV \)\\
\hline
\( M_{\rho } \)&
 \( 763\, MeV \)&
 \( 763\, MeV \)&
 \( 763\, MeV \)&
 \( 770\, MeV \)&
 \( 770\, MeV \)\\
\hline
\( m_{s} \)&
 \( 492\, MeV \)&
 \( 492\, MeV \)&
 \( 492\, MeV \)&
 \( 511.2\, MeV \)&
 \( 550\, MeV \)\\
\hline
\( g_{2}=\kappa /2 \)&
 \( 12.17\, fm^{-1} \)&
 \( 12.17\, fm^{-1} \)&
 \( 12.17\, fm^{-1} \)&
 \( 7.23\, fm^{-1} \)&
 \( 9.62\, fm^{-1} \)\\
\hline
\( g_{3}=\lambda /6 \)&
 \( -36.259 \)&
 \( -36.259 \)&
 \( -36.259 \)&
 \( 0.6183 \)&
 \( 61.24 \)\\
\hline
\( g_{N\sigma } \)=\( g_{s} \)&
 \( 10.138 \)&
 \( 10.138 \)&
 \( 10.138 \)&
 \( 10.0289 \)&
 \( 6.9667 \)\\
\hline
\( g_{\omega } \)&
 \( 13.285 \)&
 \( 13.285 \)&
 \( 13.285 \)&
 \( 12.6139 \)&
 \( 5.8561 \)\\
\hline
\( g_{\rho } \)&
 \( 8.941 \)&
 \( 9.214 \)&
 \( 10.751 \)&
 \( 9.2644 \)&
 \( 8.3803 \)\\
\hline
\( c_{3} \)&
 \( 0 \)&
 \( 0 \)&
 \( 0 \)&
 \( 71.3075 \)&
 \( 0 \)\\
\hline
\( \Lambda _{v} \)&
 \( 0 \)&
 \( 0.005 \)&
 \( 0.025 \)&
 \( 0 \)&
 \( 0 \)\\
\hline
\( g_{\omega \sigma } \)&
 \( 0 \)&
 \( 2/3\, g_{N\sigma } \)&
 \( 2/3\, g_{N\sigma } \)&
 \( 0 \)&
 \( 0.6617 \)\\
\hline
\( g_{\rho \sigma } \)&
 \( 0 \)&
 \( 2/3\, g_{N\sigma } \)&
 \( 2/3\, g_{N\sigma } \)&
 \( 0 \)&
 \( 0.6617 \) \\
\hline
\end{tabular}
\par{}
\caption{\label{tab1}Parameter sets for the Lagrangian (\ref{lag}).}
\end{table}%

\end{document}